\pdfoutput=1
\documentclass[10pt,pra,aps,twocolumn,showpacs,superscriptaddress]{revtex4}
\usepackage{amsmath}
\usepackage{latexsym}
\usepackage{amssymb}
\usepackage{graphics,epstopdf}
\usepackage{graphicx}
\usepackage[colorlinks=true]{hyperref}
\usepackage{float}

\begin{document}

\title{Multipartite Bell-type Inequality by Generalizing Wigner's Argument}

\author{Dipankar Home}
\email{dhome@jcbose.ac.in}
\affiliation{Bose Institute, CAPSS, Block-EN, Sector-V, Salt Lake City, Kolkata-700091, India}

\author{Debashis Saha}
\email{debashis7112@iiserkol.ac.in}
\affiliation{Indian Institute of Science Education and Research Kolkata, Mohanpur 741252, India}

\author{Siddhartha Das}
\email{sdas@students.iiserpune.ac.in}
\affiliation{Indian Institute of Science Education and Research Pune, Pune 411008, India}

\begin{abstract}
Wigner's argument inferring Bell-type inequality for the EPR-Bohm entangled state is generalized here for any $N$-partite state. This is based on assuming for the relevant dichotomic observables the existence of the overall joint probability distributions, satisfying the locality condition,  that would yield the measurable marginal probabilities. For any $N$, such Generalized Wigner's Inequality (GWI) is violated by quantum mechanics for all pure entangled states. The efficacy of GWI is probed, comparing with the Seevinck-Svetlichny multipartite Bell-type inequality, by calculating threshold visibilities for the quadripartite GHZ, Cluster and W states that determine their respective robustness with respect to the quantum mechanical violation of GWI in the presence of white noise. 
\end{abstract} 

\pacs{03.65.Ud, 03.67.Bg, 03.67.Mn}

\maketitle

\section{ Introduction} Closely following the discovery of Bell-CHSH inequality \cite{Bell} showing an incompatibility between quantum mechanics and local realism, Wigner \cite{Wigner} gave an interesting derivation of a different Bell-type inequality, albeit restricted for the EPR-Bohm entangled state. This was based on assuming the locality condition and existence of the joint probability distributions (JPD) for the occurrence of different possible combinations of the outcomes of measurements of the relevant observables. However, Wigner's approach has remained largely unexplored, apart from its application in the context of entangled neutral kaons \cite{Domenico,Bramon}, and its extension made by Castelleto et al. \cite{gw} for an arbitrary two qubit state in order to study its implication for quantum key distribution. Against this backdrop, the central theme of the present paper is to bring out the wider significance of Wigner's argument by extending it to develop an elegant method that yields a generalized multipartite inequality, essentially based on the assumption of the existence of JPD. The Generalized Wigner Inequality (GWI) thus obtained turns out to be useful for probing multipartite quantum nonlocality for an arbitrary $N$-partite state. Before proceeding to develop our generalization of Wigner's approach and comparing the results based on this approach with that obtained from other directions of studies concerning multipartite Bell-type inequalities, let us briefly outline the derivation of Wigner's original inequality in a way suitable for our subsequent treatment.

In the scenario considered by Wigner, two spin-$1/2$ particles are prepared in a singlet state and are spatially separated for which the spin components of the particles, respectively, are measured along three directions, say, $a,b$ and $c$. Then, in this context, considering the individual outcomes ($\pm 1$) of nine possible pairs of measurements, Wigner's original inequality can be derived as follows.

Assuming the locality condition and an underlying stochastic hidden variable (HV) distribution corresponding to a quantum state specified by a wave function, one can infer in the HV space the existence of overall joint probabilities for the individual outcomes of measuring the pertinent observables, from which the observable marginal probabilities can be obtained. In particular, the single probability of the occurrence of a particular outcome of measuring an observable for either the first or the second particle can be obtained as a marginal of the assumed overall joint probability distributions - note that, in conformity with the locality condition, this is fixed irrespective of what variable of the other particle is measured. Thus, corresponding to an underlying stochastic HV, say $\lambda$, one can define $p_\lambda(v_1(a),v_1(b),v_1(c);v_2(a),v_2(b),v_2(c))$ as the overall joint probability of occurrence of the outcomes, where $v_1(a)$ represents an outcome ($\pm 1$) of the measurement of the observable $a$ for the first particle, and so on. For example,  $p_\lambda(+,-,-;-,+,+)$ expresses the overall joint probability of occurrence of the outcomes $v_1(a)=+1,v_1(b)=-1,v_1(c)=-1$ for the first particle, and $v_2(a)=-1,v_2(b)=+1,v_2(c)=+1$ for the second particle. Then, the joint probability, say, $v_1(a)=+1$ and $v_2(b)=+1$ for the first and the second particle respectively can be written, using the perfect anti-correlation property of the singlet state, as $ p_\lambda(a+,b+)=p_\lambda(+,-,+;-,+,-)+p_\lambda(+,-,-;-,+,+)$. Similarly, writing $p_\lambda(c+,b+)$ and $p_\lambda(a+,c+)$ as marginals, and assuming non-negativity of the overall joint probability distributions in the HV space, it can be shown that
\begin{equation} \label{eq0}
p_\lambda(a+,b+)\leq p_\lambda(a+,c+)+p_\lambda(c+,b+)
\end{equation} Subsequently, by integrating over the hidden variable space for an arbitrary distribution, one can obtain the original form of Wigner's inequality
\begin{equation} \label{eq1}
p(a+,b+)\leq p(a+,c+)+p(c+,b+).
\end{equation}
where $p(a+,b+)$ is the observable joint probability of getting $+1$ for both the outcomes if the observables $a$ and $b$ are measured on the first and the second particle respectively, and so on.

If the respective angles between $a$ and $b$, $a$ and $c$, $b$ and $c$ are $\theta_{12},\theta_{13}$ and $\theta_{23}$, then substituting the QM expressions for the relevant joint probabilities in the inequality given by Eq.(\ref{eq1}) one obtains $\frac{1}{2} \sin^2(\theta_{12}/2) \leq \frac{1}{2} \sin^2(\theta_{13}/2) +\frac{1}{2} \sin^2(\theta_{23}/2)$ - a relation which is {\it not} valid for arbitrary values of $\theta_{12},\theta_{13}, \theta_{23}$. This shows an incompatibility between quantum mechanics (QM) and Wigner's form of inequality given by Eq. (\ref{eq1}), restricted for the singlet state in the bipartite case. Note that the above argument is within the framework of stochastic HV theory, subject to the locality condition, and does not depend on using the notion of determinism. Here we may stress that the incompatibility of QM with Eq. (\ref{eq1}) rules out only a class of stochastic HV theories satisfying the locality condition. This is also true for the subsequent generalization of Wigner's inequality discussed in this paper in terms of stochastic HV theory.

\section {Generalized Wigner Inequality for any bipartite state}
For the purpose of our treatment here, we consider that pairs of dichotomic observables $a$ or $a'$ and $b$ or $b'$ are measured on the first and the second particle respectively. For the generality of our treatment here, we assume, consistent with the locality condition, an underlying HV distribution given by $\rho(\lambda)$ such that for $2^4$ possible combinations of pairs of outcomes, each such pair of outcomes occur with a certain probability in the HV space. Now, defining $p_{\lambda}(v_1(a),v_1(a'); v_2(b), v_2(b'))$ to be the overall joint probability pertaining to a particular $\lambda$, the joint probability $p_{\lambda}(a+,b+)$ is assumed to be obtainable as a marginal in the HV space, given by the following expression
\begin{equation} \label{eq2}
\begin{split}
 p_{\lambda}(a+,b+)=\sum\limits_{v(a')}\sum\limits_{v(b')}p_{\lambda}(+,v_1(a');+,v_2(b'))
\\ =p_{\lambda}(+,+;+,+)+p_{\lambda}(+,+;+,-)+ \\ p_{\lambda}(+,-;+,+)+p_{\lambda}(+,-;+,-)
\end{split}
\end{equation}
Writing expressions similar to Eq.(\ref{eq2}) for the other marginal joint probabilities given by $p_{\lambda}(a+,b'+)$, $p_{\lambda}(a'+,b+)$, $p_{\lambda}(a'-,b-)$, and invoking non-negativity of the overall joint probabilities in the HV space, we obtain the following result
\begin{eqnarray}\label{eq3}
p_{\lambda}(a+,b'+)+p_{\lambda}(a'+,b+)+p_{\lambda}(a'-,b'-)\nonumber \\=p_{\lambda}(a+,b+)+8 \mbox{~non-negative~ terms}. 
\end{eqnarray}

\noindent Then, it follows that $p_{\lambda}(a+,b'+)+p_{\lambda}(a'+,b+)+p_{\lambda}(a'-,b'-)-p_{\lambda}(a+,b+)\geq0$ for any $\lambda$. Subsequently, integrating over the hidden variable space using the distribution $\rho(\lambda)$, one obtains the following form of GWI for bipartite systems
\begin{equation}\label{eq4}
\begin{split}
p(a+,b+)-p(a+,b'+)-p(a'+,b+)-p(a'-,b'-)\leq0
\end{split}
\end{equation}
where we have used $\int \rho(\lambda)d\lambda = 1$, while each term in Eq. (\ref{eq4}) is the observable joint probability.

Similarly, other forms of GWI can be derived, such as the one given below
\begin{equation}\label{eq5}
 p(a+,b-)-p(a+,b'-)-p(a'+,b-)-p(a'-,b'+)\leq0
\end{equation}
Next, substituting the QM expressions for the joint probabilities pertaining to \textit{any} state (pure or mixed) in Eqs.(\ref{eq4}) and Eq.(\ref{eq5}), one obtains in terms of the expectation values the following inequalities $\langle ab\rangle-\langle ab'\rangle-\langle a'b\rangle-\langle a'b'\rangle \leq2$ and $ \langle a'b'\rangle+\langle ab'\rangle+\langle a'b\rangle-\langle ab\rangle \leq2$  respectively,
which together imply
\begin{equation}\label{eq6}
 |\langle a'b'\rangle+\langle ab'\rangle+\langle a'b\rangle-\langle ab\rangle| \leq2
\end{equation}
It is, thus, seen that the QM violation of GWI can be regarded as equivalent to violating Bell-CHSH inequality for bipartite systems. Note that, a special case of the QM violation of GWI of the form given by Eq.(\ref{eq4}) is when $p(a+,b+)>0, p(a+,b'+)=p(a'+,b+)=p(a'-,b'-)=0$, which, interestingly, is the case considered in Hardy's non-locality argument \cite{hardy}. Another special case of GWI occurs by taking $a'=b'=c$, whence Eq.(\ref{eq4}) reduces to

\begin{equation}\label{eq7}
 p(a+,b+)-p(a+,c+)-p(c+,b+)-p(c-,c-)\leq0
\end{equation}

\noindent which was derived by Castelletto et al. \cite{gw}. 

\section {Generalization for any $N$-partite state}
The derivation of GWI for multipartite systems is a suitable extension of the procedure adopted for bipartite systems. Here we consider two dichotomic observables for each of the spatially separated particles which are denoted by $a_i$ and $a'_i$ where the index $i$ represents the $i^{th}$ particle. Following an argument similar to that used earlier for writing Eq.(\ref{eq2}), considering any $N$-partite system, one can write the marginal joint probability $p_{\lambda}(a_1+,a_2+,a_3+,...,a_N+)$ in the $\lambda$-space in terms of the overall joint probabilities as follows

\begin{equation}\label{eq9}
\begin{split}
& p_{\lambda}(a_1+,a_2+,a_3+,...,a_N+) =\\ 
& \sum\limits_{v(a'_1),v(a'_2),...v(a_N')} p_{\lambda} (+,v(a'_1);+,v(a'_2);+,v(a'_3);...;+,v(a'_N))
\end{split}
\end{equation}

\noindent Here $p_{\lambda}(v(a_1),v(a'_1); v(a_2), v(a'_2);...;v(a_N), v(a'_N))$ is defined to be the overall joint probability pertaining to a given $\lambda$ where $v(a_{i})$ denotes the possible outcomes $(\pm 1)$ when the observable $a_{i}$ is measured on the $i$-th particle and so on. Writing expressions similar to Eq. (\ref{eq9}) for the other relevant marginal joint probabilities given by $p_{\lambda}(a'_1+,a_2+,a_3+,...,a_N+),$ $p_{\lambda}(a_1+,a'_2+,a_3+,..., a_N+),$ $p_{\lambda}(a_1+,a_2+,a'_3+,...,a_N+),........$, $p_{\lambda}(a_1+,a_2+,...,a'_N+)$ and $p_{\lambda}(a'_1-,a'_2-,a'_3-,...,a'_N-)$, and by invoking non-negativity of the overall joint probabilities, it can be seen that
\begin{equation} 
\label{eq10}
\begin{split}
& p_{\lambda}(a'_1+,a_2+,a_3+,...,a_N+)+ p_{\lambda}(a_1+,a'_2+,a_3+,...,a_N+)  \\ 
& + p_{\lambda}(a_1+,a_2+,a'_3+,...,a_N+)+...+p_{\lambda}(a_1+,a_2+,...,a'_N+) \\
& + p_{\lambda}(a'_1-,a'_2-,a'_3-,...,a'_N-)= p_{\lambda}(a_1+,a_2+,a_3+,...,a_N+) \\
&+N2^{N} \mbox{non-negative terms} \\
\end{split}
\end{equation}
Therefore, we can write the following inequality
\begin{equation}\label{eq11}
\begin{split}
p_{\lambda}(a_1+,a_2+,a_3+,...,a_N+)\\-p_{\lambda}(a'_1+,a_2+,a_3+,...,a_N+)\\-  p_{\lambda}(a_1+,a'_2+,a_3+,...,a_N+)\\
-p_{\lambda}(a_1+,a_2+,a'_3+,...,a_N+)\\- ... -p_{\lambda}(a_1+,a_2+,a_3+,...,a'_N+)\\-p_{\lambda}(a'_1-,a'_2-,a'_3-,...,a'_N-)\leq0
\end{split}
\end{equation}
for each $\lambda$. Subsequently, by integrating over the HV space for an arbitrary $\rho(\lambda)$, one obtains from Eq.(\ref{eq11}), the following form of $N$-partite GWI in terms of the observable joint probabilities, given by
\begin{equation}\label{eq12}
\begin{split}
& p(a_1+,a_2+,a_3+,...,a_N+)-p(a'_1+,a_2+,a_3+,...,a_N+) - \\& p(a_1+,a'_2+,a_3+,...,a_N+)
-p(a_1+,a_2+,a'_3+,...,a_N+) \\&.......-p(a_1+,a_2+,a_3+,...,a'_N+)\\& - p(a'_1-,a'_2-,a'_3-,...,a'_N-)\leq0
\end{split}
\end{equation}
It can be checked by substituting in Eq. (\ref{eq12}) the QM expressions for the joint probabilities for \textit{any} state that the above form of GWI can be written in terms of the expectation values, as in Eq. (\ref{eq6}), so that the local realist upper bound for the QM violation of $N$-partite GWI is obtained to be `$N$'. Note that, interestingly, the form of GWI given by Eq. (\ref{eq12}) is equivalent to Cereceda's multipartite local realist inequality \cite{jose} that was obtained by generalizing Hardy's argument for quantum nonlocality. Possible implications of this equivalence that suggests a close link between Wigner's and Hardy's argument should be worth investigating. The maximum QM violation of this form of inequality was studied by Ghosh and Roy \cite{ghosh}. Recently, Yu et. al. \cite{Yu} have provided a powerful demonstration that {\it all} pure $N$-partite quantum entangled states violate such an inequality. Here it is worth emphasizing that the derivation of Eq. (\ref{eq12}) is within the framework of stochastic HV theory, without requiring to invoke the notion of determinism and satisfying the locality condition in the sense of the single probability of the occurrence of an outcome of measurement on a particular particle being unaffected by what observable is  actually  measured on any of the other particles.

At this stage, before proceeding further, to put things in historical perspective, we recall that studies related to $N$-partite local realist inequalities were initiated by Mermin \cite{Mermin} and Roy-Singh \cite{RS} in order to investigate the nature of QM violation of local realism for large $N$. Later, Seevinck-Svetlichny $N$-partite local realist correlation inequality \cite{Svetlichny} was derived using the assumption of partial factorizability. Further, a set of multipartite inequalities were derived pertaining to a local realist description of $N$-particle correlation by Zukowski and Brukner \cite{Zukowski1, Zukowski2}. Unlike for GWI, not all pure entangled states violate these inequalities. Also, in contrast to GWI, the upper local realist bound for these $N$-partite correlation inequalities increases exponentially with $N$.

We now proceed to discuss the key features of GWI, moving from the tripartite to the quadripartite case, concentrating mainly on the latter.

\section {Efficacy of GWI} The form of GWI given by Eq.(\ref{eq12}) reduces in the {\it tripartite} case to the following form
\begin{equation}\label{eq13}
\begin{split}
 p(a_1+,a_2+,a_3+)-p(a_1+,a_2+,a'_3+)-p(a_1+,a'_2+,a_3+)\\-p(a'_1+,a_2+,a_3+)-p(a'_1-,a'_2-,a'_3-)\leq0
\end{split}
\end{equation}
It is interesting that such an inequality for the tripartite case was earlier obtained from Hardy's argument of quantum nonlocality, and it was argued that QM violates this inequality for all pure tripartite entangled states \cite{Oh, Kar}.

Next, considering the {\it quadripartite} case of GWI, we obtain from Eq.(\ref{eq12}) the following inequality
\begin{equation}\label{eq14}
\begin{split}
 p(a_1+,a_2+,a_3+,a_4+)-p(a_1+,a_2+,a_3+,a'_4+) \\ -p(a_1+,a_2+,a_3'+,a_4+)
-p(a_1+,a'_2+,a_3+,a_4+) \\ -p(a'_1+,a_2+,a_3+,a_4+)-p(a'_1-,a'_2-,a'_3-,a'_4-)\leq0
\end{split}
\end{equation}
 which, in terms of the expectation values, reduces to
\begin{equation}\label{eq15}
\begin{split}
\langle S \rangle =  \langle a_1a_2a_3a_4\rangle - \langle a'_1a'_2a'_3a'_4\rangle - \langle a_1a_2a_3a'_4\rangle - \langle a_1a_2a'_3a_4\rangle \\ - \langle a_1a'_2a_3a_4\rangle - \langle a'_1a_2a_3a_4\rangle - \langle a_3a'_4\rangle - \langle a_2a'_4\rangle - \langle a_1a'_4\rangle \\ - \langle a'_3a_4\rangle - \langle a_2a'_3\rangle - \langle a_1a'_3\rangle -
\langle a'_2a_3\rangle - \langle a'_2a_4\rangle - \langle a_1a'_2\rangle \\ - \langle a'_1a_3\rangle - \langle a'_1a_4\rangle - \langle a'_1a_2\rangle - \langle a_1a_2\rangle - \langle a_1a_3\rangle - \langle a_1a_4\rangle \\ -
\langle a_3a_4\rangle - \langle a_2a_3\rangle - \langle a_2a_4\rangle - \langle a'_3a'_4\rangle -\langle a'_2a'_3\rangle -
 \langle a'_2a'_4\rangle \\ - \langle a'_1a'_3\rangle - \langle a'_1a'_4\rangle - \langle a'_1a'_2\rangle-
\langle a_2a_3a'_4\rangle- \langle a_1a_3a'_4\rangle \\ - \langle a_1a_2a'_4\rangle- \langle a_2a'_3a_4\rangle-\langle a_1a'_3a_4\rangle - \langle a_1a_2a'_3\rangle- \langle a'_2a_3a_4\rangle \\ - \langle a_1a'_2a_3\rangle-\langle a_1a'_2a_4\rangle-  \langle a'_1a_3a_4\rangle-
\langle a'_1a_2a_3\rangle- \langle a'_1a_2a_4\rangle \\ + \langle a'_2a'_3a'_4\rangle+\langle a'_1a'_3a'_4\rangle+\langle a'_1a'_2a'_4\rangle+
\langle a'_1a'_2a'_3\rangle \\ -2(\langle a_1\rangle+\langle a_2\rangle+\langle a_3\rangle+\langle a_4\rangle)\leq4
\end{split}
\end{equation}

Now, since the GHZ state \cite{GHZ},  Cluster state \cite{Cluster} and the W state \cite{W} are of special interest in the context of quantum information, here we consider these types of entangled states. In the quadripartite case, the expressions for these states are respectively

\begin{subequations}
\begin{equation}
\label{wl1}
|GHZ\rangle_{1234} = \frac{1}{\sqrt{2}}(|0000\rangle + |1111\rangle)
\end{equation}
\begin{equation}
\label{wl2}
|Cluster\rangle_{1234} = \frac{1}{2}(|0000\rangle + |0011\rangle + |1100\rangle - |1111\rangle)
\end{equation}
\begin{equation}
\label{wl3}
|W\rangle_{1234} = \frac{1}{2}(|0001\rangle + |0010\rangle + |0100\rangle + |1000\rangle)
\end{equation}

\end{subequations}

\noindent where $|0\rangle$ and $|1\rangle$ are the eigenstates of $\sigma_{z}$ corresponding to the eigenvalues $+1$ and $-1$ respectively.

In order to calculate the maximum QM violation of GWI in the quadripartite case corresponding to the GHZ state, we consider the settings in the $X-Y$ plane given by 
\begin{equation}\label{o1}
\begin{split}
a_i = \cos(\phi_i) \sigma_x + \sin(\phi_i) \sigma_y\\ a'_i = \cos(\phi'_i) \sigma_x + \sin(\phi'_i) \sigma_y
\end{split}
\end{equation} \noindent where $\phi_{i, i^{\prime}}$ is the angle with the $X$ axis. 
For the GHZ state (\ref{wl1}), it can be shown that all the correlation functions occurring on the LHS of (15) involving the four observables of the type considered in Eq.(\ref{o1}) are $cosine$ of the sum of the corresponding angles with $X$ axis, and the correlation functions  involving less than four observables are all zero. Then the LHS of GWI in the quadripartite case given by Eq.(\ref{eq15}) reduces to the following form
\begin{equation}\label{ghz}
\begin{split}
\langle S \rangle_{GHZ} = \cos(\phi_1+\phi_2+\phi_3+\phi_4) - \cos(\phi'_1+\phi'_2+\phi'_3+\phi'_4) \\ - \cos(\phi_1+\phi_2+\phi_3+\phi'_4) -
\cos(\phi_1+\phi_2+\phi'_3+\phi_4)\\ - \cos(\phi_1+\phi'_2+\phi_3+\phi_4) - \cos(\phi'_1+\phi_2+\phi_3+\phi_4)
\end{split}
\end{equation} 
Now, simplifying Eq. (\ref{ghz}) by choosing $\phi_1+\phi_2+\phi_3+\phi_4 = \alpha$ and $\phi'_i = \phi_i + \beta$ we obtain
\begin{equation}
\langle S \rangle_{GHZ} = \cos(\alpha) - \cos (\alpha + 4\beta) - 4\cos(\alpha+ \beta)
\end{equation}
The maximum value of the above quantity signifying the maximum QM violation of GWI for the GHZ state in the quadripartite case is given by $ 5.656848 (\approx 4\sqrt{2})$ when $(\alpha, \beta) = (0.6981, 2.2427)~or~(5.5938, 4.0492)$ in radian.

Next, in order to calculate the maximum QM violation for the quadripartite Cluster and W states, we consider the settings in the $X-Z$ plane given by 
\begin{equation}
\begin{split}
a_i = \cos(\phi_i) \sigma_z + \sin(\phi_i) \sigma_x\\ a'_i = \cos(\phi'_i) \sigma_z + \sin(\phi'_i) \sigma_x
\end{split}
\end{equation} where $\phi_{i, i^{\prime}}$ is the angle with the $Z$ axis. 
Considering these observables, the expression for the LHS of  GWI in the quadripartite case given by Eq.(\ref{eq15}) for the Cluster state (\ref{wl2}) is given by
\begin{widetext}
\begin{equation}\label{cluster}
\begin{split}
& \langle S \rangle_{Cluster} = cos(\phi'_3)sin(\phi'_1)sin(\phi'_2) - cos(\phi_1)cos(\phi'_2) - cos(\phi'_1)cos(\phi_2) - cos(\phi'_1)cos(\phi'_2) -  cos(\phi_3)cos(\phi_4) \\& - cos(\phi_3)cos(\phi'_4) - cos(\phi'_3)cos(\phi_4) - cos(\phi'_3)cos(\phi'_4) - cos(\phi_3)sin(\phi_1)sin(\phi'_2) - cos(\phi_3)sin(\phi'_1)sin(\phi_2) \\& - cos(\phi'_3)sin(\phi_1)sin(\phi_2) - 
cos(\phi_1)cos(\phi_2) - cos(\phi_4)sin(\phi_1)sin(\phi'_2) - cos(\phi_4)sin(\phi'_1)sin(\phi_2) -  cos(\phi'_4)sin(\phi_1)sin(\phi_2) \\& + cos(\phi'_4)sin(\phi'_1)sin(\phi'_2) - cos(\phi_1)sin(\phi_3)sin(\phi'_4) -  cos(\phi_1)sin(\phi'_3)sin(\phi_4) - cos(\phi'_1)sin(\phi_3)sin(\phi_4) \\& + cos(\phi'_1)sin(\phi'_3)sin(\phi'_4) -  cos(\phi_2)sin(\phi_3)sin(\phi'_4) - cos(\phi_2)sin(\phi'_3)sin(\phi_4) - cos(\phi'_2)sin(\phi_3)sin(\phi_4) \\& +  cos(\phi'_2)sin(\phi'_3)sin(\phi'_4)  + cos(\phi_1)cos(\phi_2)cos(\phi_3)cos(\phi_4)  -  cos(\phi_1)cos(\phi_2)cos(\phi_3)cos(\phi'_4) - cos(\phi_1)cos(\phi_2)cos(\phi'_3)cos(\phi_4) \\& - cos(\phi_1)cos(\phi'_2)cos(\phi_3)cos(\phi_4)  - cos(\phi'_1)cos(\phi_2)cos(\phi_3)cos(\phi_4) -  cos(\phi'_1)cos(\phi'_2)cos(\phi'_3)cos(\phi'_4)
\end{split}
\end{equation}
\end{widetext}
By choosing $\phi_1 = - \phi_2 = - \phi_3 = \phi_4$, $\phi'_1 = - \phi'_3$, $ \phi'_2 = - \phi'_4,$ and $ \phi'_1 + \phi'_2 = 2\pi$, the expression given in Eq. (\ref{cluster}) reduces to 

\begin{equation}
\begin{split}
& \langle S \rangle_{Cluster} = cos^4(\phi_1) - cos^4(\phi'_1) - 2cos^2(\phi'_1) - 2cos^2(\phi_1)\\& + 4cos^3(\phi'_1) - 4cos^2(\phi_1)cos(\phi'_1)  - 4cos^3(\phi_1)cos(\phi'_1)\\& - 4cos(\phi_1)cos(\phi'_1) + 8cos(\phi_1)sin(\phi_1)sin(\phi'_1)
\end{split}
\end{equation}
Now, for $(\phi_1, \phi'_1) = (0.3578, 2.2689) ~or~ ( 5.9341, 4.0230)$ in radian, the value of the above expression is 5.7442. This is indeed the maximum QM violation of GWI given by Eq. (15) for the quadripartite Cluster state that is confirmed by the numerical study.

Considering the quadripartite W state given by Eq. (\ref{wl3}), the LHS of GWI given by Eq. (\ref{eq15}), while choosing $\phi_1 = 0,\phi_2 = \phi_4, \phi'_2 = \phi'_4$, reduces to the following expression
\begin{widetext}
\begin{equation}\label{w}
\begin{split}
& \langle S \rangle_{W} = cos(2\phi_2)/4 + cos(2\phi'_2)/4 - cos(2\phi_2 - \phi_3)/8 + cos(2\phi_2 - \phi'_3)/8 + cos(\phi'_1 + 2\phi_2 + \phi_3)/2 + cos(\phi'_1 + 2\phi'_2 + \phi'_3)/2 \\& + cos(\phi_2 + \phi'_2 - \phi_3)/4 + cos(\phi_2 - \phi'_2 + \phi_3)/4 + cos(\phi'_1 + \phi_2)/2 + cos(\phi'_1 + \phi'_2)/2 + cos(\phi'_1 + \phi_3)/2 + (3cos(\phi_2 + \phi'_2))/2 \\& + cos(\phi'_1 + \phi'_3)/2 + cos(\phi_2 + \phi_3)/2 + (3cos(\phi_2 + \phi'_3))/2 + (3cos(\phi'_2 + \phi_3))/2  + cos(\phi'_2 + \phi'_3)/2 + cos(\phi'_1 - 2\phi_2 - \phi_3)/8 \\& + cos(\phi'_1 + 2\phi_2 - \phi_3)/8 + cos(\phi'_1 - 2\phi'_2 - \phi'_3)/8 + cos(\phi'_1 + 2\phi'_2 - \phi'_3)/8 + cos(\phi_2 - \phi'_2 - \phi_3)/4 - 2cos(\phi_2) - (5cos(\phi_3))/4 \\& + cos(\phi'_3)/4 - cos(\phi'_1 - \phi_2)/2 + cos(\phi'_1 + 2\phi_2)/2  - cos(\phi'_1 - \phi'_2)/2 - cos(\phi'_1 + 2\phi'_2)/2 - cos(\phi'_1 - \phi_3)/4 - cos(\phi_2 - \phi'_2)/2 \\& - cos(\phi'_1 - \phi'_3)/4 - cos(\phi_2 - \phi_3)/2 - cos(\phi_2 - \phi'_3)/2 - cos(\phi'_2 - \phi_3)/2 - (5cos(2\phi_2 + \phi_3))/8 - cos(\phi'_2 - \phi'_3)/2 \\& + (9cos(2\phi_2 + \phi'_3))/8 - cos(2\phi'_2 + \phi'_3)/2  + cos(\phi'_1 + \phi_2 + \phi_3) - cos(\phi'_1 + \phi'_2 + \phi'_3) + (9cos(\phi_2 + \phi'_2 + \phi_3))/4 - 3/2
\end{split}
\end{equation}
\end{widetext}
For $(\phi'_1,\phi_2,\phi'_2,\phi_3,\phi'_3)$=$(2.271,0.131,2.298,-2.557,-0.892)$ in radian, the value of the expression given by Eq. (\ref{w}) is 6.5603, which is  confirmed to be the maximum value by the numerical investigation.

Thus, the maximum QM violations of GWI of the form given by Eq.(\ref{eq15}) that have been obtained for the quadripartite GHZ, Cluster and W states correspond to the (approximate) values on the LHS of Eq. (\ref{eq15}) given by $5.6568$, $5.7442$ and $6.5603$ respectively. Note that, for GWI, the Cluster and W states show greater QM violation than the GHZ state, whereas in the case of the Seevinck-Svetlichny inequality (SSI), the GHZ state shows greater (maximal) QM violation than the Cluster and W states. Next, using these states, in order to illustrate the efficacy of GWI, we probe their tolerance to white noise with respect to the QM violation of GWI. For this, let us introduce the notion of what is known as the visibility parameter pertaining to a state. Considering a quadripartite mixed state given by 

\begin{equation}
\label{eq16}
\rho = v|\psi\rangle \langle \psi| + (1-v)\frac{I}{2^{4}}
\end{equation}

\noindent where $|\psi\rangle$ is a quadripartite pure state and the parameter $v$ is defined as the visibility of the state $|\psi\rangle$. Note that, $(1-v)$ denotes the amount of white noise present in the state $\rho$, while for $v=0$, $\rho$ denotes maximally mixed state. The minimum value of $v$ for which the QM predictions for $\rho$ given by Eq. (\ref{eq16}) violate a given local realist inequality, therefore, signifies the maximum amount of white noise that can be present in the state $\rho$ for the persistence of the QM violation of the given local realist inequality. This value of $v$ is known as the threshold visibility pertaining to the state $|\psi \rangle$ corresponding to the given inequality. 

The threshold visibilities for the GHZ, Cluster and W states, corresponding to the N-partite GWI given by Eq. (\ref{eq12}) are found to be 0.7071 $(\approx \frac{4}{4\sqrt{2}})$, 0.6964 $(\approx \frac{4}{5.7442})$ and 0.6097 $(\approx \frac{4}{6.5603})$ respectively, using the feature that in the presence of white noise, for any state, the QM expression of the LHS of Eq.(\ref{eq15}) is $v$ times the expression without noise. On the other hand, using SSI, the threshold visibility is found to be minimum $(\sim 0.7071)$ for the GHZ state \cite{Svetlichny}. Thus, an interesting feature is that, in the quadripartite case, the threshold visibility using the GHZ state for GWI turns out to be the same as the threshold visibility for this state corresponding to SSI, while for the Cluster and W states, the threshold visibilities for GWI are less than that corresponding to SSI. This means that the QM violations of GWI for the quadripartite Cluster and W states can persist in the presence of greater amount of white noise than that in the case of SSI.

\section {Conclusion}
Here we may stress that although there have been different directions of studies exploring multipartite Bell-type inequalities, our paper provides a distinct approach through N-partite generalization of Wigner's argument for obtaining Bell-type inequality. In particular, it will be interesting to investigate if there is any relation between the QM violation of GWI and the various measures of entanglement of quantum states \cite{Geometric}. Curiously, for quadripartite systems, the increasing order of pure entangled states, namely, GHZ, Cluster and W states, considered on the basis of the magnitudes of the maximum QM violation of GWI shown by these states, turns out to be the same as the ordering of these states when considered on the basis of the magnitudes of the measure of the `persistency of entanglement' \cite{Cluster} for these states. Its possible implications need to be investigated in future studies.

Finally, we note that stimulating questions about the nature of quantum nonlocality have been raised in the light of studies showing that a suitably constructed classical model using an appropriate two-particle phase space distribution can be employed to show the violation of Bell-CHSH inequality; further, the role of the factorizability condition used in deriving the Bell-CHSH inequality for stochastic hidden variables has been subjected to a critical examination \cite{Alex}. On the other hand, Fine \cite{Fine} had shown that the assumption of the existence of JPD and the use of the factorizabilty condition are equivalent in the study of quantum nonlocality for bipartite systems. Our work, based on GWI, serves to validate the notion that, assuming the locality condition and the existence of overall joint probabilities in any stochastic HV theory yielding the measurable marginal probabilities is sufficient to demonstrate, for the multipartite states, an incompatibility between QM and a class of stochastic HV theories satisfying the locality condition. Possible ramifications of such findings regarding fundamental questions about the nature of quantum nonlocality for multipartite systems call for comprehensive studies.

\begin{acknowledgements}
D. H. thanks Alex Matzkin for helpful comments. S. D. thanks Sibasish Ghosh for useful discussions. We also thank the Referees for their thoughtful suggestions. This work is supported by the Department of Science and Technology, Government of India. D. H., D. S. and S. D. thank the Centre for Science, Kolkata, IISER-Kolkata, and IISER-Pune respectively for support.
\end{acknowledgements}


\begin{thebibliography}{99}
\bibitem{Bell} J. S. Bell, Physics \textbf{1}, 195 (1964); J. F. Clauser, M. A. Horne, A. Shimony and R A Holt, Phys. Rev. Lett. \textbf{23}, 880 (1969).
\bibitem{Wigner} E. P. Wigner, Am. J. Phys. \textbf{38}, 1005 (1970).
\bibitem{Domenico} A. D. Domenico, Nucl. Phys. \textbf{B 450}, 293 (1995).
\bibitem{Bramon} A. Bramon and M. Nowakowski, Phys. Rev. Lett. \textbf{83}, 1 (1999).
\bibitem{gw} S. Castelletto, I. P. Degiovanni and M L Rastello,, Phys. Rev. A \textbf{67}, 044303 (2003).
\bibitem{hardy} L. Hardy, Phys. Rev. Lett. \textbf{71}, 1665 (1993).
\bibitem{jose} J. L. Cereceda, Phys. Lett. A \textbf{327}, 433 (2004).
\bibitem{ghosh} S. Ghosh and S. M. Roy, J. Math. Phys. \textbf{51}, 122204 (2010).
\bibitem{Yu} S. Yu, Q Chen, C. Zhang, C. H. Lai, and C. H. Oh, Phys. Rev. Lett. \textbf{109}, 120402 (2012).
\bibitem{Mermin} N. D. Mermin, Phys. Rev. Lett. \textbf{65}, 1838 (1990).
\bibitem{RS}S. M. Roy and V. Singh, Phys. Rev. Lett. \textbf{67}, 2761 (1991).
\bibitem{Svetlichny} M. Seevinck and G. Svetlichny, Phys. Rev. Lett. \textbf{89}, 060401 (2002).
\bibitem{Zukowski1} M. Zukowski and C. Brukner, Phys. Rev. Lett. \textbf{88}, 210401 (2002).
\bibitem{Zukowski2} M. Zukowski, C. Brukner, W. Laskowski and M. Wiesniak, Phys. Rev. Lett. \textbf{88}, 210402 (2002).
\bibitem{Oh} J. L. Chen, C. Wu, L. C. Kwek and C. H. Oh, Phys. Rev. A \textbf{78}, 032107 (2008).
\bibitem{Kar} S. K. Choudhary, S Ghosh, G. Kar and R. Rahaman, Phys. Rev. A \textbf{81}, 042107 (2010).
\bibitem{GHZ} D. M. Greenberger, M. A. Horne and A. Zeilinger: Bell's theorem, Quantum Theory, and Conceptions of the Universe, Kluwer Academics, Dordrecht, The Netherlands (1989), pp. 73-76.
\bibitem{Cluster} H. J. Briegel and R. Raussendorf, Phys. Rev. Lett. \textbf{86}, 910 (2001).
\bibitem{W} W. Dur, G. Vidal, J. I. Cirac, Phys. Rev. A \textbf{62} 062314 (2000).
\bibitem{Geometric} M. Aulbach, D. Markham and M. Murao, New J. Phys. \textbf{12}, 073025 (2010); S. Tamaryan, T. C. Wei, D. K. Park, Phys. Rev. A \textbf{80}, 052315 (2009).
\bibitem{Alex} A. Matzkin, Phys. Rev. A \textbf{77}, 062110 (2008); J. Phys. A: Math. Theor. \textbf{41}, 085303 (2008); M. Golshani and A. Fahmi, Ann. de la Found. L. de Broglie \textbf{26}, 735 (2001).
\bibitem{Fine} A. Fine, Phys. Rev. Lett. \textbf{48}, 291 (1982).


\end{thebibliography}
\end{document}